\documentclass{elsarticle}
\usepackage{color}
\usepackage{comment}
\usepackage{amsmath}
\usepackage[htt]{hyphenat}
\usepackage{natbib}

\begin{document}
%

\title{Student Cluster Competition 2017, Team University of Texas at Austin/Texas State University: Reproducing Vectorization of the Tersoff Multi-Body Potential on the Intel Skylake and NVIDIA V100 Architectures}
%
%
%
%
%


\author[ut]{James Sullivan}
\ead{jsull3@utexas.edu}
\author[tacc]{Collin Weir}
\ead{cweir@tacc.utexas.edu}
\author[ut]{Austin Reichert}
\ead{austinskyreichert@yahoo.com}
\author[tacc]{R. Todd Evans}
\ead{rtevans@tacc.utexas.edu}
\author[tacc]{W. Cyrus Proctor}
\ead{cproctor@tacc.utexas.edu}
\author[tacc]{Nicolas Thorne}
\ead{nthorne@tacc.utexas.edu}

\address[ut]{The University of Texas at Austin, 110 Inner Campus Drive, Austin, Texas 78705}
\address[tacc]{Texas Advanced Computing Center, 10100 Burnet Rd, Austin, Texas, 78758}
\maketitle
\begin{abstract}
This paper satisfies the reproducibility challenge of the Student Cluster Competition at SC17\footnote{\label{SCC17}http://www.studentclustercompetition.us}. We attempt to reproduce the results of a vectorized code for the Tersoff multi-body potential kernel of the molecular dynamics code Large-scale Atomic/Molecular Massively Parallel Simulator (LAMMPS). In addition, we outline a series of performance measurements across a number of different processing architectures. We investigated accuracy, optimization, performance, and scaling with our Intel CPU and NVIDIA GPU based cluster.
\end{abstract}


\section{Introduction}
Reproducibility is a trait all scientific papers should aspire toward, and work in the high-performance community is no exception. There is still significant room for improvement, and the importance of the issue of reproducibility has brought us to consider the paper from SC16 titled ``The Vectorization of the Tersoff Multi-Body Potential: An Exercise in Performance Portability'' \citep{hohnerbach:tersoff}. In this paper, Hohnerbach et al.~applied a vectorization scheme to the computationally intense Tersoff multi-body potential, which requires calculations that move beyond pairwise interactions, relying instead on local complex interactions involving relative positions between multiple particles. The authors attempt to optimize the Tersoff kernel for LAMMPS \cite{LAMMPS} and increase its portability to various cluster architectures. The authors provide an implementation in their code to accomplish both these goals through vectorization and related methods.

The authors' code provides for three vectorization schemes for appropriate architectures, and optimizes both the USER-INTEL and USER-KOKKOS packages. The implementation of their vectorization scheme splits their algorithm into computational and filter parts. The filtering forces LAMMPS to wait to compute until the vector lanes are full or near-full, increasing the number of operations performed in a given time interval. The authors make the following claims: 
\begin{itemize}
    \item their single and mixed precision code additions are accurate when run compared to double precision calculations,
    \item their code increases performance for running calculations of the Tersoff potential across CPU-only, GPU co-processor, and Xeon Phi co-processor systems,
    \item and that their vectorized code scales well across various hardware configurations.
\end{itemize}
 
We investigated all of these claims using some of the most powerful CPU and GPU hardware options available at the time of this writing. This gave us the ability to analyze and determine how well the claims of vectorization portability have held up on contemporary hardware. Moreover, we also applied the Tersoff potential to the task of calculating the thermodynamic properties of silicon \citep{porter}. This ``porter'' problem potentially provides an additional test for the vectorization scheme of Hohnerbach et al., the portability of the scheme to other systems, and the reproducibility of the authors' results. 

This paper is organized as follows: we describe our cluster in terms of hardware and software and our build process for the application in Section 2, we present our attempt to reproduce the results of the original paper in Section 3, and we conclude in Section 4.

\section{Cluster and Builds}
\subsection{Machine description}

The machine used was an eight-node cluster consisting of eight Dell PowerEdge R740s. Each node contained two 24-core 2.1GHz Intel Xeon Platinum 8160 ``Skylake'' processors amounting to 48 cores per node. Additionally, each node contained one NVIDIA Tesla V100 GPU, and had a total 192GB DRAM. The cluster ran CentOS 7.4, was connected by a Mellanox InfiniBand EDR switch capable of transmitting 100Gb/s, and had high-performance storage managed with two striped Intel P4500 Series 4.0TB NVMe SSDs.

The competition required us to source hardware from sponsors and then operate our cluster below 3000 Watts while competing. The original paper involved runs on various compute architectures, including CPU-only runs, GPU accelerated runs, and Intel Xeon Phi co-processor accelerated runs. Due to our hardware constraints, we only attempted to replicate the CPU-only runs and GPU accelerated runs. These runs were sufficient to satisfy the terms of the competition.

\subsection{Compilation/Run description}

Using the March 2016 version of LAMMPS pulled directly from the \texttt{lammps-tersoff-vector} repository of the original paper, we compiled different builds of LAMMPS with and without the vectorization scheme of Hohnerbach et al.~using the USER-INTEL, USER-KOKKOS, and USER-GPU packages. For the USER-INTEL package, we built using the Intel MPI (17.0.4) \cite{impi}, Math Kernel \cite{mkl} and Threading Building Blocks \cite{tbb} libraries, and each package was compiled with the Intel C++ Compiler (17.0.4) \cite{ic++}. We followed the procedure of Hohnerbach et al.~for building with the USER-KOKKOS package and used the Kokkos package from the Trilinos library \cite{KOKKOS}. Additionally, we used CUDA 9.0 \cite{CUDA} to build both the USER-KOKKOS package and USER-GPU packages. We also used OpenMP for both the USER-INTEL and USER-GPU builds.

To properly build with the USER-GPU package, we used the Tesla70 CUDA architecture specification corresponding to the V100s. However, for the USER-KOKKOS package associated with the repository containing the code of Hohnerbach et al., the most recent available CUDA architecture was Maxwell53. There was no version of Kokkos yet released for the Tesla70 architecture at the time of this work. This meant that our USER-KOKKOS build will likely experience increased performance with an updated version of Kokkos corresponding to the Tesla70 architecture. As in Hohnerbach et al., we also use the most advanced vector instructions available for a given processor architecture, in our case those enabled by the ``skylake-avx512'' flag, in all of our builds.

\section{Results}

\subsection{Accuracy study}

\begin{figure}
    \centering
    \includegraphics[width=0.9\textwidth]{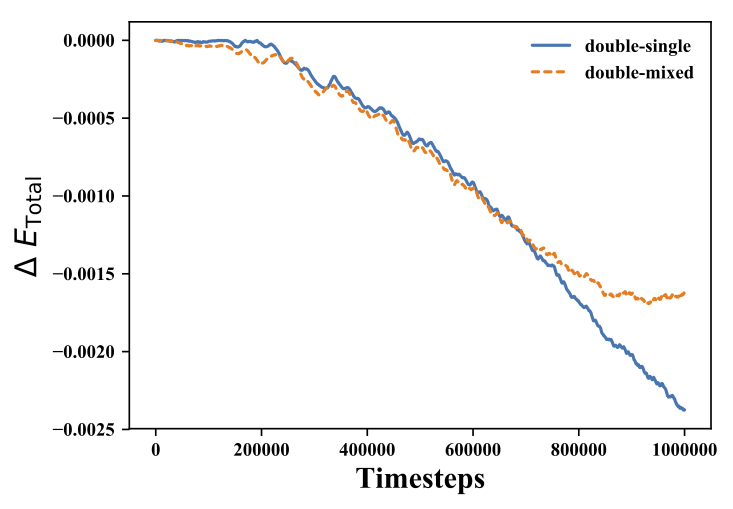}
    \caption{Accuracy study of LAMMPS precision level. Evaluation of single and mixed precision implementations when compared to double precision. Relative difference in values of total energy produced by both the single vs.~double (solid blue) and mixed vs.~double precision solvers (dashed orange) for 32,000 atoms for $10^{6}$ steps.} 
    \label{fig:accuracy}
\end{figure}

Hohnerbach et al.~claim that the single and mixed precision versions of their code are accurate when compared with the double precision offered by LAMMPS by default. Their evidence for this claim is in a comparison of relative difference between single and double precision in the total energy of a long-running simulation. The authors quote a relative difference of 0.002\% in this total energy for 32,000 atoms over $10^{6}$ timesteps. We test this claim on our cluster, and go a step further by comparing the relative difference between their mixed and double precision implementations as well.

In an attempt to replicate the accuracy of Hohnerbach et al., we calculated the relative difference in total energy using the same problem size and number of timesteps as in the original paper (Fig.~\ref{fig:accuracy}). We find a relatively low maximum value of deviation between the double and reduced precision runs of less than 0.25\%. This is much larger than the deviation quoted in Hohnerbach et al., but is still small, so we verify their claim of accuracy but to a weaker degree. Notably, the deviation grows nearly monotonically after about $2.5 \times 10^{5}$ timesteps for the single precision case.  The mixed-precision case's deviation initially grows at the same rate as the single-precision case but appears to be bounded after roughly $9 \times 10^{5}$ timesteps. Both cases' behavior is different from the random but non-increasing deviation displayed in Fig.~3 of the original paper. These discrepant behaviors may be due to a variety of factors including differences in instruction set and/or optimizations performed at compile time.  

\subsection{Performance differences}

\begin{figure}
    \centering
    \includegraphics[width=0.9\textwidth]{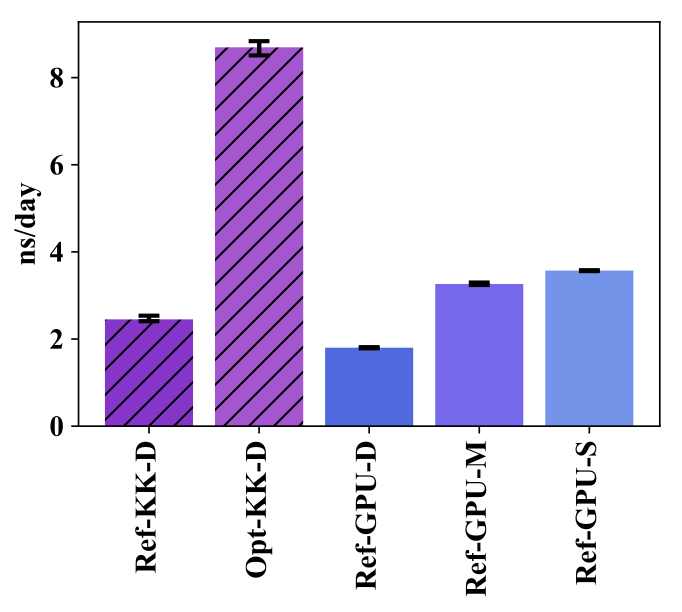}
    \caption{Evaluation of performance portability for NVIDIA V100s for 512,000 atoms. Kokkos runs are shown in purple (hatched) and reference GPU runs are in blue (unhatched). Error bars are shown with limits denoting the minimum and maximum measured ns/day over 5 runs. Ref-KK-D uses the non-optimized USER-KOKKOS package, Opt-KK-D uses the optimized USER-KOKKOS package, Ref-GPU-D, Ref-GPU-M, and Ref-GPU-S use the  USER-GPU package with double, mixed, and single precision, respectively. }
    \label{fig:gpu}
\end{figure}

In this Section, we replicate nearly exactly the results of Hohnerbach et al.~using GPU offloading in the LAMMPS calculations. The only difference lies in our problem size. Whereas the authors originally used a specific GPU benchmarking problem with 256,000 atoms, we solve a similar problem with 512,000 atoms using the same number of timesteps (2420). Here Opt-KK refers to calculations run using the optimized code with the USER-KOKKOS package, while Ref-KK refers to non-optimized calculations using this package. Ref-GPU refers to calculations using the non-optimized USER-GPU package, and is run with single, mixed, and double precision. 

We find results (Fig.~\ref{fig:gpu}) that are very similar to those of the original paper, and verify the claim of increased performance on GPUs. Even at a problem double in size, there is about a factor of two increase in performance (measured in ns/day) between single and double precision runs, but only a very slight growth in performance between mixed and single precision runs. Without the optimization applied, Ref-KK performs better than the similar double precision REF-GPU-D, but performs worse compared to the mixed and single precision GPU runs. 

The most significant result we find is the relatively large increase in performance for Opt-KK-D. When compared to Ref-KK-D we find nearly a factor of 4X increase in performance. This is only a slightly greater factor than the one found for the same setup in Hohnerbach et al., but the near 3X increase in performance between Opt-KK-D and the Ref-GPU-M/S cases is an improvement upon the authors' original increase of 1.5X in performance in these cases. This is likely due to the closer balance of single precision to double precision capability found in an NVIDIA V100 (2 to 1) versus a NVIDIA K20x or K40 ($\sim 3$ to 1).

It is also interesting that the absolute ns/day values we quote for our GPU runs are much higher than those offered by the authors in their original work. This is likely due to the architectural improvements on the NVIDIA V100 as well as the updated CUDA Toolkit \cite{CUDA}. In the original paper, the authors' highest value in ns/day for any GPU offloaded run was accomplished with Opt-KK-D on a NVIDIA K40 GPU with about 1.8 ns/day. In comparison, when running on the NVIDIA V100s, the lowest value we find is at about the same value of 1.8 ns/day for Ref-GPU-D. The highest-performing run (Opt-KK-D) on the V100s gives a value of approximately 8.5 ns/day, almost 5X the value produced for the same run on the K40. Even without using Kokkos, it would seem that upgrading GPU hardware would outperform the best possible outcome of optimization. However, the surprisingly high return on performance in a large dataset using Opt-KK-D illustrates the power of the vectorization scheme of Hohnerbach et al. For future use, note that building using Kokkos limits the CUDA compute capability to version 5.3, while the V100s are capable of 7.0. Thus, we might expect an even greater performance gain if Kokkos could use this updated architecture.

We also attempted to run the porter problem while offloading to GPU, but were unsuccessful. The reason for this is two-fold. First, Kokkos in the version of LAMMPS provided by the original authors does not support the minimize function required to run the porter problem. Second, we were not able to run the non-optimized reference versions built using the GPU package, since we received an error due to the inability of the GPU package to use a triclinic crystal system box required by the porter problem.

\subsection{Scaling study}

\begin{figure}
    \centering
    \includegraphics[width=0.9\textwidth]{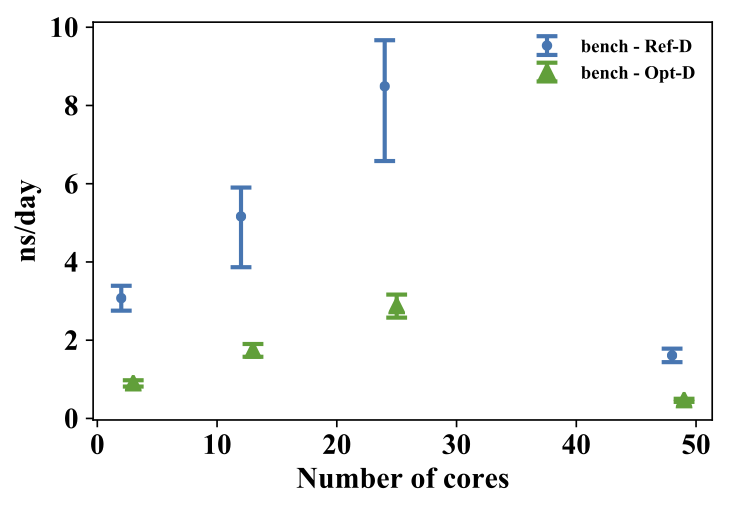}
    \includegraphics[width=0.9\textwidth]{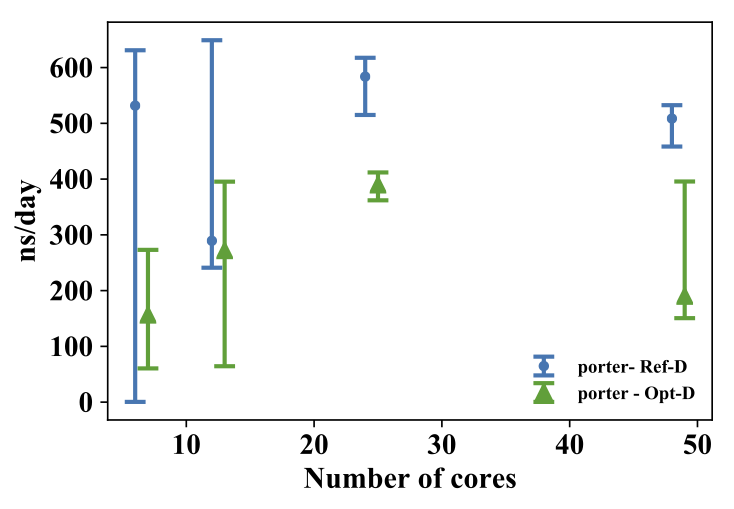}
    \caption{Optimization results with scaling across one node with two 24-core Skylake sockets (Intel Platinum 8160) for 512,000 atoms. \textit{Top panel}: strong scaling of tersoff problem runs. \textit{Bottom panel}: strong scaling of porter problem runs. The error bars represent the maxima and minima of reported ns/day for our 5 runs.}
    \label{fig:scale1}
\end{figure}

\begin{figure}
    \centering
    \includegraphics[width=0.9\textwidth]{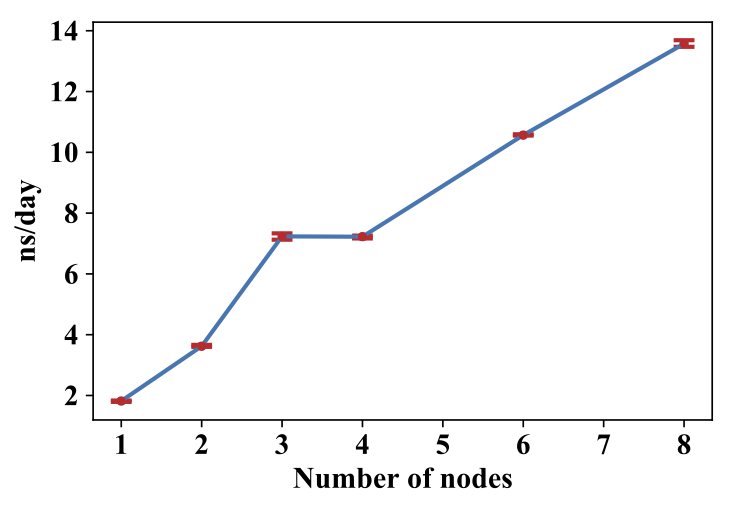}
    \caption{Reference GPU (using the USER-GPU package) results with strong scaling across eight nodes with NVIDIA V100s for 512,000 atoms using double precision. We see roughly linear scaling across our cluster. The error bars represent the maxima and minima of reported ns/day for our 5 runs.}
    \label{fig:scale2}
\end{figure}

Hohnerbach et al.~originally perform a scaling study on nodes both with and without Xeon Phi Knight's Corner coprocessors to support their claim of scalability for large problems. For the competition, we performed a similar study for nodes with GPUs, as well as an additional study of scaling within a node. We turn to the latter first. We used our Intel Skylake processors possessing 24 cores each, more than any socket available to the original authors, to do so. We also did this in part due to our lack of access to the Xeon Phi co-processors, though further investigation using those would certainly be worthwhile. 

We performed a run with a problem size of 512,000 atoms utilizing a varied number of cores with no explicit affinity settings. We found an intra-socket speedup of 8X when scaling from 1 core to 24 cores. This scaling, however, did not continue when using cores from both sockets as can be seen in Fig.~\ref{fig:scale1}.  In fact, using all 48 cores on a node resulted in comparable or worse performance than using a single core.  Interestingly, we saw a smaller speedup in the porter problem, but not such a large drop-off in performance at a large number of cores. In both cases, we saw a rise in performance across one socket, and a drop as we used the whole node, though with looser time constraints we would dedicate more time to getting more data in the region between 12 and 48 cores. It is also important to keep in mind the size of the error bars on these reported values, especially for the porter problem where we see large variability in ns/day when running with a small number of cores. Perhaps the size of the variability is due to the nature of the porter problem itself, but it would be worth repeating runs for more than just the 5 we completed to better understand this behavior.  Again, competition time constraints prevented us from generating further scaling data.
Our results imply scaling within a socket is excellent and beyond a single socket poor for these problem sizes.  However, the performance gains from using the optimized code is retained at all scales much as in the original paper.

To see if the scaling exhibited by the Xeon Phis is also present for scaling with GPUs, we performed an additional scaling study using our V100 GPUs.
In Fig.~\ref{fig:scale2}, we present the scaling in performance measured as we increased the number of GPUs in our runs using the USER-GPU package. The error bars are very small here, showing consistency in our runs. We used 512,000 atoms, in order to provide a direct comparison to the CPU results that were gathered from the competition. This was a deviation from the 256,000 atom size presented in the original paper so we could demonstrate the increased performance of the GPUs when compared to the results generated from the CPU runs. We observe roughly linear scaling for the double precision run that was used as a reference in the original paper on our V100 GPUs, which is superior to the original reported scaling. Thus, we support the original claim of scaling across nodes with accelerators. 

Interestingly, there is super-linear scaling observed when going from 2 to 3 nodes as seen in Fig~\ref{fig:scale2}. There is the possibility that the problem size could fit inside the memory of 3 GPUs but not 2.  This could reduce data transfer overhead. The resulting benefit to performance could then be overwhelmed by communication overhead when using 4 or greater nodes. A second possible explanation is that the MPI configuration happened to be tuned to perform well using 3 nodes.  Unfortunately, without access to the original hardware we can't be sure of the cause.  

We were unable to perform the GPU-based scaling study using the Kokkos build and associated optimizations. The LAMMPS package documentation for USER-KOKKOS states Kokkos cannot run on multiple GPUs, but if Kokkos were to provide multi-GPU support in the future, such a run would provide an interesting followup to this large-scale scaling study.

\section{Conclusions}

We attempted to replicate the results of Hohnerbach et al.~on a GPU-enabled cluster. We found using CPU tests of accuracy that the optimization is still accurate, though there is a nearly monotonic decline in accuracy of the single precision versions at large timesteps. In the future it is worth considering the effect on accuracy of Xeon Phi co-processor acceleration versus CPU only or GPU accuracy. However, this decline is small and though a trend is present, we replicate the order of accuracy produced in Hohnerbach et al.

When examining performance differences using GPUs, we found significant performance increases due to our V100 GPUs. We found similar relative performance in the various precision and optimization cases for GPU offloading. As a result, we were able to replicate this outcome. In the future, we suggest testing hardware such as the V100s along with its appropriate CUDA architecture, i.e.~a more recent version of compute capability, for increased performance. 

We test scaling on a 48-core Skylake node and find good scaling up to 24 cores and poor scaling when using all 48 cores in the node.  This is in contradiction to what is observed in Hohnerbach et al., where scaling from 24 to 48 cores is efficient.  It is not a direct comparison, however; the scaling presented in the original paper is for whole nodes, going from 1 24-core node to 2 24-core nodes, while the scaling in our study was restricted to within a node.

In a more direct comparison to the original paper's scaling results using accelerators, we found excellent scaling on NVIDIA V100 GPUs across many nodes for the reference GPU run. In fact, the scaling we observed was better than that reported for the Phi's. We thus replicated, and even exceeded, the strong increase in performance across multiple nodes for accelerators reported by Hohnerbach et al. In the future, it may be interesting to consider scaling on multi-GPU runs.

\section{Acknowledgments}
We would like to thank the other members of our Student Cluster Competition Team, Vivian Nguyen, Jeremie Gallegos, Carlos Hurtado, as well as Cyrus Proctor, Nick Thorne, and Todd Evans for extensive mentoring and support. We also thank our corporate sponsors, including TACC, Dell, NVIDIA, Intel, Geist, Mellanox, and Raytheon. We thank the Student Cluster committee for their hard work organizing and running the competition, especially Kris Garrett for being patient with our questions.

\section{References}
%
\bibliographystyle{abbrv}
\bibliography{refs}  
%

\end{document}